# EL MOVIMIENTO DE LAS SOMBRAS

## Una propuesta de trabajo para la escuela secundaria

**Alejandro Gangui**
*Instituto de Astronomía y Física del Espacio, UBA-CONICET*
**María C. Iglesias y Cynthia P. Quinteros**
*Centro de Formación e Investigación en la Enseñanza de las Ciencias, FCEyN-UBA*

Numerosas investigaciones extranjeras y algunas locales ponen de manifiesto las dificultades que encuentran los alumnos de la escuela media para comprender algunos fenómenos astronómicos. Ello nos llevó a reflexionar sobre el estado de la enseñanza y el aprendizaje de esos temas. Entre las dificultades, podemos mencionar la de reconocer los cambios en los aspectos observables del movimiento del Sol: duración del día, lugares y momentos de la salida y puesta del Sol, su altura máxima en el cielo, etcétera.

Además, muy pocos estudiantes son capaces de identificar los momentos singulares del año (los equinoccios y los solsticios) y las regularidades que acaecen en torno a ellos, lo cual va asociado con una visión distorsionada de cómo se producen los cambios astronómicos a lo largo del año. Tampoco suelen reconocer la existencia de representaciones o modelos alternativos que puedan dar cuenta de algunas observaciones, ni se hace uso operativo en los colegios de las hipótesis de los modelos para explicar observaciones conocidas. Como regla general, los alumnos no utilizan hipótesis para explicar fenómenos relacionados, aun si se les pregunta explícitamente por un hecho que no pueden justificar.

Formadores e investigadores en la didáctica de la astronomía señalan que la recurrencia de las llamadas *ideas previas* se prolonga más allá de la escuela secundaria y afecta también a estudiantes de magisterio y a profesores en actividad. Frente a estas cuestiones, cabe preguntarse qué podemos hacer los docentes universitarios e investigadores para superar los obstáculos que impiden o dificultan la construcción del conocimiento científico. Como plantea, por ejemplo, Martínez Sebastià, la enseñanza de la astronomía no puede limitarse a saber que la Tierra es una esfera que gira sobre sí misma y alrededor del Sol: también debe poder contribuir a explicar los fenómenos astronómicos mediante dichas hipótesis.

Creemos que un buen camino para paliar estas deficiencias es la enseñanza-aprendizaje por investigación, que supone enfrentar a los alumnos con diversas preguntas o situaciones problemáticas. De la misma forma que en la ciencia los conocimientos se elaboran en respuesta a preguntas, este enfoque considera que una enseñanza basada en situaciones problemáticas favorece un aprendizaje significativo.

A la luz de lo dicho, presentamos aquí una propuesta de trabajo para ser llevada al aula por los docentes. Tiene una pregunta o problema como hilo conductor, a modo de guía, que da sentido a la secuencia presentada. La pregunta es: *¿cómo utilizar las sombras para construir un reloj que marque las horas? S*e espera que, al intentar responder, los alumnos aborden algunos temas de astronomía de manera progresiva y secuenciada, partiendo de sus conocimientos previos y contrastando los temas estudiados con ellos. Al finalizar, podrán construir un reloj de Sol y explicar su funcionamiento.

Si bien muchos de los principales contenidos sobre los que se trabaja en esta secuencia se corresponden con los del 9º año de la Educación General Básica de la Argentina (*La Tierra, el universo y sus* *cambios* de los núcleos de aprendizaje prioritarios), algunas jurisdicciones incluyen determinados de ellos en otros momentos de la escuela secundaria. Por ejemplo, en el primer año de la escuela secundaria de la provincia de Buenos Aires se trata, bajo el título *La Tierra y el universo*, el tema *Los objetos del sistema solar y sus movimientos*.

### Material de trabajo

Para llevar adelante esta propuesta se requiere el siguiente material:

* Cuatro esferas de telgopor (tres de 4cm de diámetro y una de 12cm de diámetro).
* Algunos escarbadientes, para proyectar sombras sobre las esferas de telgopor.
* Una plancha de telgopor de unos 40cm x 40cm aproximadamente, para constituir la base de una maqueta.
* Dos varillas delgadas de madera: una de 10cm de largo, para ser utilizada como gnomon de la maqueta, y otra de unos 16cm de largo, para servir de eje de rotación terrestre. (Llamamos *gnomon* al indicador de las horas de los relojes de sol.)
* Un poco de plastilina, para fijar la varilla en posición vertical.
* Una linterna, que, emulando al Sol, permitirá crear las sombras.
* Una lámpara incandescente de por lo menos 60 watts, con su portalámparas.
* Tres alambres semirrigidos de 1mm de diámetro y unos 80cm de longitud cada uno, aproximadamente, para colocar en la maqueta y representar los arcos diurnos solares en tres momentos del año.
* Un estuche o caja de plástico rígido de discos compactos (CD o DVD), para formar la estructura rígida de un reloj de Sol.
* Unos 15cm de hilo de algodón, para usar como gnomon del reloj.
* Cinta engomada y un par de ganchos mariposa.

### Secuencia de actividades

#### ACTIVIDAD 1
Perdí mi reloj y no se qué hora es. ¿Cómo podría saber la hora si desaparecieran todos los relojes?

*El objetivo de esta actividad es plantear a los alumnos el tema de trabajo. Es una indagación introductoria que los lleva, guiados por el docente, a reflexionar sobre algunos fenómenos cotidianos relacionados con el movimiento aparente del Sol y la producción de sombras. En particular, les permitirá analizar el enigma de cómo establecer las horas del día sin recurrir a un reloj.*

Para dar inicio a la actividad sugerimos algunas preguntas que guíen la discusión acerca de los fenómenos que pretendemos considerar con los alumnos:

* ¿Cómo hacían para saber la hora nuestros antepasados, cuando no existían relojes?
* ¿Qué objeto en la naturaleza cambia en forma contínua y es observable durante el día? ¿De qué manera lo hace? Escriban en una hoja sus propuestas. Descríban el objeto elegido.
* Si pierdo mi reloj, ¿de qué manera podría saber la hora?

Mientras transcurre la discusión, es muy probable que aparezca en los alumnos la idea recurrente sobre los puntos de salida y puesta del Sol. Es frecuente oirles afirmar que dichos fenómenos ocurren *exactamente* en el este y en el oeste. Por lo tanto, se sugiere registrar todas sus respuestas en un cartel o afiche de construcción colectiva, para volver sobre ellas más adelante. Por otro lado, los estudiantes secundarios ya suelen tener incorporada la noción de las sombras que arroja el Sol, y es probable que en la discusión anterior haya surgido ese tema. Entre las preguntas que orienten la discusión, sugerimos entonces las siguientes acerca de las sombras y su relación con el movimiento aparente del Sol.

* ¿Qué conocen ustedes sobre las sombras que arroja el Sol? Durante el transcurso del día, la sombra de un objeto cualquiera, ¿apunta siempre en la misma dirección? ¿Tiene siempre el mismo tamaño?
* En el caso de que las sombras no sean iguales, ¿a qué se debe que su dirección cambie y lo mismo su tamaño? ¿Por qué sucede esto?
* Si esta misma actividad la realizáramos en otros momentos del año, ¿piensan ustedes que veríamos cambios notorios en las sombras? ¿Cuáles? ¿Por qué?

Muchas de las preguntas sugeridas abren la discusión sobre los cambios de la trayectoria diurna del Sol con las estaciones del año. En el transcurso de la discusión puede resultar útil apelar al recurso de la modelización: con una linterna y una varilla de madera (nuestro gnomon) mantenida en forma vertical sobre una hoja blanca con la ayuda de un poco de plastilina se pueden representar el arco diurno solar, los diferentes momentos del día, las características de las diferentes estaciones, etcétera. Lo importante es lograr que los alumnos se imaginen o representen el movimiento del Sol a lo largo del día y del año.

Si bien los estudiantes han tenido oportunidades de trabajar estos temas en años anteriores, puede ocurrir que no tengan clara la relación entre trayectoria del Sol y producción de sombras. O que ignoren de qué manera varían los arcos solares en el transcurso del año. También es frecuente que, a pesar de haber trabajado estos temas, consideren que el Sol se encuentra exactamente sobre nuestras cabezas (en el cenit) en el mediodía solar, algo que solo puede suceder en localidades ubicadas entre los trópicos. Si ese fuera el caso, sugerimos destinar un tiempo importante a realizar observaciones y registros al aire libre, y analizar con los alumnos cómo deberían ser las sombras si las posiciones del Sol o su movimiento fueran como ellos suponen.

Para hacer lo anterior, conviene definir, entre todos, qué aspectos registrar. Por ejemplo, la posición del Sol y la orientación de la sombra, el momento del día en que se realiza la observación, el día del año. También sugerimos utilizar el modelo constituido por linterna y gnomon, que permite representar sombras imposibles de observar y registrar con el Sol de determinato momento y ubicación geográfica.

Algunas de las preguntas apropiadas para esta instancia podrían ser:

* Si quisiéramos registrar las sombras producidas por el Sol ¿qué tipo de objeto deberíamos utilizar? (Se supone que los alumnos tomaron conciencia en años escolares anteriores que los objetos opacos producen sombra. Sugerimos recuperar ese concepto).
* ¿Qué sería importante registrar en el momento de realizar nuestras observaciones?
* ¿Podremos registrar todas las sombras en el transcurso del día? ¿Cómo podríamos estudiar aquellas de las cuales no poseemos datos?

Al finalizar esta primera actividad los alumnos habrán tenido oportunidad de acercarse a algunos conocimientos sobre la trayectoria aparente del Sol. Por ejemplo, el significado de levante y poniente (así como el de las expresiones cotidianas *el Sol sale* o *el Sol se pone*), el movimiento aparente del Sol durante el día (arco diurno solar), los arcos solares en las diferentes estaciones del año, la hora solar y su diferencia con la hora civil, el mediodía solar, etcétera.

Si bien lo considerado hasta el momento puede resultar conocido por los alumnos, es importante refrescar esos conocimientos para poder pasar a la identificación de los días singulares (equinoccios y solsticios) y de las regularidades en torno a ellos.

Una pregunta o situación problemática que conduce a la siguiente actividad (la actividad 2) y que pretende aproximar a los alumnos a cómo se producen los cambios astronómicos a lo largo del año, es:

* Si pudieramos registrar todas las sombras producidas por el Sol durante un año, ¿encontraríamos cada día sombras distintas? ¿Podríamos encontrar sombras iguales en diferentes días del año? ¿Cuándo y porqué? No olvidar de escribir las respuestas en el cartel mencionado antes.

Sabemos que el arco solar va cambiando gradualmente de día en día. Es corto en invierno (y por ello los días son cortos) y largo, igual que la duración de los días, en verano. De hecho, el arco solar se va alargando –y elevando en el cielo– desde el solsticio de invierno hasta el solsticio de verano. El mediodía solar corresponde al momento en el que el Sol se halla justo en el medio de su arco. Por lo tanto, a mediodía el Sol se hallará más alto en verano que en invierno.

La palabra solsticio viene del latin y significa *sol quieto*, pues los cambios en su altura sobre el horizonte a mediodía culminan –es decir, alcanzan su máximo o su mínimo– y *cambian de sentido* ese día. Por lo tanto, si el Sol venía constantemente aumentando (o disminuyendo) su altura y luego pasa a disminuirla (o aumentarla), tiene que haber un instante en el que invierte el sentido del cambio, y en el que, por ello, está quieto, como le sucede a un vehículo que pasa de avanzar a retroceder (o viceversa).

Si se mide la longitud de las sombras a mediodía, resultan crecientes día a día durante seis meses del año, y decrecientes durante los otros seis. Esto quiere decir, además, que las sombras que proyectan los objetos, digamos, cuatro días antes de un solsticio serán aproximadamente iguales a las que proyectan cuatro días después. No hay forma de distinguir entre estos dos instantes del año solo mediante la medición de la longitud de las sombras. Sin comprender la regularidad de estos fenómenos astronómicos y las particularidades de los momentos singulares del año solar, es difícil comprender cabalmente la repetición de las medidas de las sombras. A eso apuntan las últimas preguntas planteadas.

También sabemos que, en todos los lugares de la Tierra, durante seis meses del año los días son más largos que las noches, y que son más cortos durante los otros seis, es decir, hay un semestre de días de más de doce horas de luz solar, y un semestre de menos de doce horas de luz. Las duraciones del día y la noche cambian gradualmente día a día, hasta alcanzar un máximo (o mínimo). Por ello, forzosamente habrá dos días al año en los que la cantidad de horas del día y la noche serán iguales. Estos son los *equinoccios* (del latín, *igual noche*). Como vimos con el caso de los solsticios, aquí nuevamente encontramos una completa simetría entre las sombras producidas por un gnomon en ambos equinoccios.

**ACTIVIDAD 2**
Diseño y construcción de una maqueta para representar el movimiento aparente del Sol durante el día.

*En esta actividad se utilizan los conceptos adquiridos en la anterior y se ponen en juego para el diseño y armado de una maqueta que represente el movimiento del Sol durante el día en distintos momentos del año. Asimismo, la actividad permite a los alumnos reflexionar sobre lo aprendido hasta el momento, a la vez que se sirve de la maqueta para comenzar a imaginar los cambios astronómicos que se suceden a lo largo del año.*

Para dar inicio a esta actividad el docente puede comentar a los alumnos que, juntos, construirán una maqueta para representar en forma simplificada el movimiento del Sol a lo largo del año. Algunas preguntas para guiar el diseño y construcción podrían ser:

* ¿Con qué materiales podríamos representar el Sol y su arco diurno aparente sobre la bóveda del cielo? La bóveda del cielo se llama *bóveda celeste*, ¿por qué? De noche, ¿forman las estrellas

una bóveda celeste? Y de día, ¿el cielo es como una bóveda? ¿Es por ello que el Sol se desplaza a lo largo de un arco? El adjetivo celeste, ¿indica en primer lugar al cielo o a su color? Si fuera lo segundo, no se podría hablar de bóveda celeste para referirse al cielo nocturno. ¿Y por qué los astros se denominan cuerpos celestes? (Evidentemente, hay muchas cosas para discutir con los alumnos.)
* ¿Qué objeto podemos colocar sobre la base de la maqueta para ver las sombras?

Es posible que los alumnos sugieran una variedad de materiales u objetos para representar el movimiento diurno del Sol. Antes de presentarles los enumerados al comienzo, el docente podrá dedicar un momento para discutir si lo sugerido por los alumnos es viable y cuán complicado resulta de implementar. Luego, con los elementos indicados, el docente guiará a los estudiantes en el armado del dispositivo. Podrá continuar esta actividad formulando ciertas preguntas o haciendo sugerencias útiles para fomentar la discusión:

* Den al alambre la forma de una semicircunferencia e insértenlo sobre una base de telgopor; luego coloquen la esferita-Sol en el punto más alto del alambre. ¿Qué representa ese punto? ¿Se relaciona con algo discutido durante la primera actividad?
* Reemplacen la esferita por una linterna y exploren las sombras a lo largo de día.
* Sabemos que la duración de los días no es la misma a lo largo del año. ¿Cómo podemos representar estos cambios en nuestra maqueta? Discútanlo entre todos y propongan agregados a la maqueta.

Luego de estas preguntas es importante alentar a los alumnos a que fundamenten sus propuestas haciendo referencia a los conocimientos (adquiridos en la actividad 1) sobre los cambios en la longitud de las sombras y las diferentes duraciones de los días a lo largo del año. Esos cambios solo pueden explicarse mediante la maqueta si modificamos los arcos solares.

Seguramente surjirá la necesidad de emplear más de un alambre o arco solar para representar la trayectoria del Sol en distintas épocas del año: en el solsticio de invierno un arco corto, en el de verano un arco largo y en los equinoccios uno intermedio (figura 1). Es interesante que el docente permita a los alumnos proponer la inclinación a dar a los arcos solares. Diferentes grupos de alumnos quizás elijan distintas inclinaciones, lo que permitirá un interesante ejercicio de reflexión colectiva. Otras preguntas útiles son:

* Utilicen la linterna para explorar las sombras a lo largo del año y para los tres arcos solares. ¿Cómo es la duración de los días? (Para facilitar el trabajo con la maqueta se puede poner un arco por vez y cambiarlo para representar los distintos momentos del año.)
* Vuelvan a mirar lo afirmado en el cartel sobre días con sombras iguales. ¿Cambiarían algo de lo dicho?
* ¿Por qué decimos que en el año hay días especiales? ¿Cuáles son esos días y qué particularidades presentan?
* Dibujen la secuencia esperada de salidas y puestas del Sol para el transcurso del año. Fundamenten sus propuestas.
* Vuelvan a colocar en la maqueta los tres alambres con las tres esferitas. Representen, en cada uno, las ocho de la mañana. ¿A qué altura deben estar, aproximadamente, las esferitas en cada alambre?
* Infieran la longitud de la sombra proyectada por el gnomon (varilla) para cada caso. Pueden utilizar la linterna para corroborar sus suposiciones.
* ¿Qué les parece que esta maqueta puede representar y qué no puede representar? ¿Cuáles son las limitaciones de la maqueta? ¿Qué modificaciones sugerirían para mejorar el dispositivo?

En la maqueta, las direcciones sobre el horizonte por donde sale o se pone el Sol (que no son otras que los puntos de inserción del alambre en la base de telgopor) están fijas. Esto no es cierto para el Sol real, ya que jamás sale o se pone en la misma posición del horizonte en que lo hizo el día anterior. Por lo tanto, se hace necesario colocar, por lo menos, tres alambres diferentes para representar los días especiales mencionados (uno para cada solsticio y uno para los dos equinoccios), como lo muestra la figura 1.

Nótese que los puntos de inserción del alambre correspondiente a los equinoccios coinciden precisamente con los dos puntos cardinales generalmente asociados con la salida y puesta del Sol. En otras palabras, el Sol solo sale exactamente por el este durante los equinoccios, y solo se pone exactamente por el oeste esos mismos días. Durante la primavera y el verano, lo hace más hacia el sur (en el hemisferio austral) y por ello el arco solar es largo; durante el otoño y el invierno lo hace más hacia el norte (nuevamente, para observadores del hemisferio sur) y por ello dicho arco es corto.

Es importante destinar un momento a trabajar con los alumnos el hecho de que el modelo o maqueta permite representar solo algunos aspectos de la realidad, como las distintas posiciones del Sol a lo largo del año o las sombras generadas por los objetos, y no otros, como las distancias relativas entre el Sol y la Tierra, o los tamaños relativos de ambos cuerpos celestes. También hay que tener en cuenta que cada grupo de alumnos puede inicialmente colocar los alambres con diferentes inclinaciones, es decir, pueden no respetar el principio de que los arcos deben ser paralelos. La inclinación de los arcos solares en función de la latitud geográfica no es objeto de discusión durante esta actividad, pero lo será en la actividad 4. Si los alumnos plantean algún interrogante al respecto, el docente puede aprovechar la discusión y preguntarles si les parece correcto representar arcos con diferentes inclinaciones y por qué. Las respuestas, anotadas en el cartel colectivo, podrán revisarse una vez finalizada la actividad 4, a la luz de los nuevos aprendizajes.

**ACTIVIDAD 3**
Representaciones o modelos teóricos usados a lo largo de la historia. Debate.

*Con esta actividad se pretende que los alumnos den cuenta de que los mismos fenómenos astronómicos se pueden explicar usando diferentes representaciones o modelos conceptuales. Una referencia a la historia de la ciencia puede enriquecer el trabajo sobre estos aspectos.*

Hasta aquí los alumnos han considerado el movimiento aparente del Sol durante el día, así como las diferencias encontradas en las distintas estaciones del año. Ahora se hace necesario hallar explicaciones que justifiquen esas constataciones. Para comenzar, se pueden plantear las siguientes preguntas:

* ¿De qué manera podemos justificar tanto el ciclo día-noche como las distintas estaciones del año?
* ¿Cómo deben ser los movimientos del Sol y de la Tierra para que ocurran estos ciclos?

Esto abrirá una instancia de debate sobre el tema. Puede ocurrir que muchos de los alumnos conozcan el movimiento de traslación de la Tierra alrededor del Sol, aunque, cuando se los interrogue con alguna profundidad, no adviertan que para describirlo se colocan fuera de la Tierra, es decir, adoptan una representación o modelo teórico distinto del que usan para hablar de la salida o puesta del Sol, cuando razonan en términos del movimiento del Sol alrededor de una Tierra fija. En esas circunstancias, conviene leer a toda la clase algunos textos que den cuenta de las diferentes cosmovisiones que los científicos han sostenido a lo largo de la historia. No es necesario, aún, que se concluya cuál imágen del mundo o modelo del cosmos es científicamente más válida; se trata simplemente de presentar modelos racionales alternativos, sobre todo dos: el geocéntrico (el universo visto desde la Tierra) y el heliocéntrico (el universo visto desde el espacio o desde el Sol). Al finalizar la lectura, se puede invitar a los alumnos a recrear dichos modelos teóricos y analizar qué aspectos puede explicar cada uno de ellos, establecer si presentan limitaciones y determinar cuál de ambos modelos ofrece las mejores explicaciones de los fenómenos estudiados (por ejemplo, el ciclo día-noche y las estaciones; se puede recurrir a CIENCIA HOY 106, p.58). Llegado el momento de la discusión grupal, para la recreación de los diferentes modelos, se podrá usar la esfera de telgopor que representa a la Tierra, la varilla de 16cm que representa su eje y la fuente de luz (lámpara incandescente). También se sugiere que los textos elegidos solo definan algunos elementos o hipótesis de cada modelo teórico, de manera tal que los alumnos deban realizar sus aportes mientras los analizan y discuten.

Durante el trabajo con las esferas de telgopor, puede ocurrir que, en un primer momento, los alumnos coloquen el eje de rotación de la Tierra en forma perpendicular al plano de la traslación de esta (la eclíptica), es decir el plano del movimiento anual de nuestro planeta alrededor del Sol. De ser así, resultaría imposible explicar las diferentes estaciones del año. La actividad permite mostrar que un modelo conceptual (la Tierra con eje inclinado respecto a la perpendicular a la ecliptica) ofrece mejores explicaciones que otro (una Tierra con eje sin inclinación). Además, permitirá apreciar el recorrido aparente del Sol para observadores ubicados en diferentes hemisferios. Este concepto será utilizado en el momento de debatir y comprender hacia dónde los alumnos deberán orientar el reloj de Sol que construirán en la actividad 5.

Después de la discusión anterior se propone una instancia de sistematización de la información, en la que los alumnos puedan elaborar un pequeño documento que dé cuenta de la dinámica de la ciencia y de cómo se va creando conocimiento. Póngase especial énfasis en las hipótesis que conforman los distintos modelos teóricos, y en los fenómenos que puede explicar cada uno de esos modelos.

Una vez finalizada la etapa de sistematización, el docente puede sugerir a los alumnos que realicen un debate sobre los dos modelos teóricos mencionados. Ideas sobre cómo organizar ese debate, aunque para otro tema, pueden encontrarse en CIENCIA HOY 106, p.36. Cada grupo tendría asignado uno de los modelos, conformándose así, por ejemplo, un debate entre *los seguidores del geocentrismo* y *los seguidores del heliocentrismo*. Los grupos deberán confeccionar una lista con los argumentos, explicaciones y evidencias que permitan sostener su postura. Asimismo, deberán contemplar y anticipar los posibles argumentos contrarios, para refutarlos.

### ACTIVIDAD 4
Las sombras y su relación con la ubicación geográfica.

*En esta actividad se propone que los alumnos utilicen los modelos teóricos geocéntrico y heliocéntrico para predecir cómo será el comportamiento de las sombras en diferentes ubicaciones de la superficie de la Tierra.*

Para comenzar, se sugiere preguntar a los alumnos cómo suponen que serán las sombras para observadores ubicados en latitudes diferentes. Es importante alentar a los alumnos a que sustenten sus explicaciones y reflexiones tomando como referencia las hipótesis de los modelos considerados en la actividad anterior. El docente puede iniciar la actividad presentando la siguiente situación:

* ¿Cómo creen ustedes que serán las sombras proyectadas por los objetos en diferentes latitudes de la superficie terrestre? ¿Cómo lo explicarían teniendo en cuenta los modelos teóricos discutidos durante la actividad 3?

Es probable que a los alumnos les resulte más simple analizar y explicar el comportamiento de las sombras en diferentes latitudes empleando la esfera de telgopor y la fuente de luz, como lo vería un observador ubicado fuera de la Tierra. A eso llamamos un *modelo externo*. Para dar inicio, los alumnos deberán primero identificar la latitud de su ciudad y luego, con un pedacito de escarbadiente, señalarla en la esfera de telgopor que hace las veces de planeta Tierra. Este palillo representará el gnomón. Para facilitar el trabajo es aconsejable señalar, en dicha esfera, la línea del ecuador. Luego, conviene pedirles que exploren las sombras producidas por el Sol (lamparita) e identifiquen diferentes momentos del día (o del año). Podrán realizar el mismo procedimiento para otras latitudes, y comparar los resultados (la longitud de las sombras, su inclinación, etcétera) con los que habían supuesto al responder a las preguntas. Asimismo, este es momento oportuno para volver sobre algunas respuestas anteriores acerca del movimiento aparente del Sol, y retornar a la noción de que, al mediodía solar, fuera de la franja tropical del planeta, el Sol jamás se encuentra en el cenit, exactamente arriba de nuestras cabezas.

Una vez que los alumnos hayan estudiado las sombras para las diferentes latitudes con el modelo externo, se hace necesario revisar la maqueta construida en la actividad 2 y, a su vez, comprender cuál es la representación que ella encierra. Para ello el docente sugerirá que los alumnos traten de pasar del modelo con el que ellos ya trabajaron (el externo, con la esfera de telgopor y la lamparita) hacia un nuevo modelo, correspondiente a un observador interno. Llamaremos a este *modelo interno (o vivencial)*. La figura 3 ilustra el trabajo por hacer: la transición lenta y reflexiva, bajo la guía del docente pero llevada a cabo por los propios alumnos, desde una representación a la otra. El modelo interno final de esta transición es, por supuesto, el mismo que los alumnos ya emplearon al trabajar con la maqueta, y que permite representar todos los fenómenos estudiados.

Sobre la base de la reciente discusión, el docente podrá sugerir las siguientes preguntas:

* Representen, con la ayuda de la maqueta y una linterna, el movimiento aparente del Sol a lo largo del año, para dos latitudes diferentes. Registren las diferencias.
* ¿Cómo son, comparados entre sí, los arcos solares para una misma latitud a lo largo del año? Revisen la maqueta construida en la actividad 2 y analicen si cumple con los requisitos que ahora señalan. Si no lo hace, realicen las modificaciones necesarias.
* Encuentren latitudes sobre la Tierra donde el Sol alcance el cenit en algún momento del año.

No es necesario enfatizar la importancia del modelo interno para la enseñanza de muchos conceptos observacionales de la astronomía, pues con ellos los observadores representan el movimiento del Sol alrededor de su ubicación geográfica sobre la superficie de la Tierra. Si bien entendemos hoy que se trata de un movimiento aparente, ello no impide que siga siendo una buena representación del ciclo día-noche y de las estaciones del año, siempre desde el punto de la observación geocéntrica. Algunas preguntas o sugerencias adecuadas para reflexionar sobre estas cuestiones son:

* Utilicemos las esferitas engarzadas en los alambres de la maqueta para tratar de definir como son, vistos desde la Tierra (*desde adentro*), los arcos solares durante los sosticios de verano y de invierno. ¿Es cierto que durante los mediodías próximos al solsticio de verano el Sol se encuentra más arriba sobre el horizonte que durante los mediodías invernales?
* Elaboren un escrito final para explicar el fenómeno de las sombras y su relación con la latitud utilizando ambos modelos teóricos.

Finalizada la actividad con modelos externos e internos, los alumnos habrán tenido la oportunidad de acercarse a la idea de que los arcos solares deben encontrarse más o menos inclinados según sea, respectivamente, mayor o menor la latitud, pero que, para una misma latitud, resultan paralelos entre sí. Por otro lado, también podrán comprender que las diferencias entre los hemisferios norte y sur radican en la dirección y el sentido en el que se mueven las sombras. En particular, notemos que durante el día las sombras de un gnomon giran a favor de las agujas del reloj *solo* en latitudes ubicadas al norte del Ecuador (fuera de los trópicos). A propósito de esta observación se puede formular a los estudiantes la pregunta que sigue:

* ¿Dónde creen que se inventaron los relojes de agujas y cuál creen que fue el modelo que estos primeros relojeros tenían en mente?

**Actividad 5**
Construcción y orientación de un reloj de Sol horizontal

*El propósito de esta actividad es que los alumnos apliquen sus conocimientos para construir su propio instrumento de medición. Se sugiere dedicar unos momentos a una explicación colectiva sobre cómo armarlo y, de paso, a refrescar los conceptos necesarios para comprender su diseño y funcionamiento.*

Para dar comienzo a esta actividad, es conveniente repasar el camino andado hasta el momento y reflexionar, junto con los alumnos, sobre cómo medir el tiempo sin un reloj convencional. Esto lleva a pensar en la construcción de un dispositivo que tenga en cuenta el movimiento de las sombras a lo largo del día. Asimismo, es importante recordar que las sombras producidas por los objetos tienen relación directa con la ubicación geográfica, dato que, por lo tanto, deberá contemplarse durante la construcción del dispositivo.

Los relojes de Sol consisten, en esencia, en un objeto que hace sombra, llamado *gnomon*, ubicado en una superficie sobre la que se proyecta dicha sombra, llamada el *tablero* del reloj. Sobre el tablero se trazan unas líneas, llamadas *líneas horarias*. El tablero es por lo común plano, pero podría también ser una superficie curva, como una esfera o un cilindro. El objeto que produce la sombra es generalmente una varilla o un hilo tenso, o a veces el borde de un cuerpo, muy comúnmente un plano triangular o escuadra y hasta el cuerpo de una persona (en el caso de algunos de los relojes llamados *analemáticos*).

Los relojes se diseñan teniendo en cuenta el hemisferio y, más precisamente, la latitud del lugar donde serán empleados: latitudes positivas corresponden al hemisferio norte; latitudes negativas corresponden al hemisferio sur. Asimismo, debemos recordar que el arco solar se inclina hacia el norte para los habitantes del hemisferio sur (y lo inverso para observadores del otro hemisferio). El Sol solo pasa por nuestro cenit en algún momento del año si nos encontramos en las regiones tropicales de la Tierra (o sea, aproximadamente entre las latitudes –23,5º y +23,5º, correspondientes a los trópicos de Capricornio y de Cáncer, respectivamente). De esta manera, para utilizar el reloj de Sol, quienes viven en el hemisferio sur, por ejemplo, en la Argentina, deben orientarlo de modo que la luz solar le llegue desde el norte.

También debemos recordar que los relojes de Sol indican la hora solar y no la hora civil. En particular, el mediodía señalado por un reloj de Sol corresponderá al mediodía solar, pero la hora civil puede ser distinta. Conviene también trabajar este tema con los alumnos.

Dado que se propone construir un reloj de Sol horizontal (figura 4), para finalizar esta actividad se puede abrir la discusión sobre otros modelos de relojes, como el mencionado analemático, o como el ecuatorial, que es uno de los más simples. Los alumnos podrán buscar información y explicar su funcionamiento, como modos de aplicar lo trabajado a lo largo de esta secuencia didáctica. El tablero de nuestro reloj de Sol horizontal se aprecia en la figura 5; fue confeccionado para la ciudad de Buenos Aires. Por supuesto, sirve también para toda otra localidad que comparta aproximadamente la misma latitud (Malargüe, en Mendoza, por ejemplo, o Ciudad del Cabo, en Sudáfrica). Reglas muy sencillas permiten confeccionar tableros de relojes utilizables en distintas latitudes de la superficie terrestre.

El dispositivo propuesto tiene la ventaja de ser útil para trabajar con los alumnos porque se trata de un modelo bastante preciso, simple de armar y cuyo funcionamiento es comprensible luego de unos pocos minutos de reflexión. Por supuesto, siempre es posible hacer dispositivos más exactos, pero en ese caso habría que incluir conceptos como la llamada *ecuación del tiempo* (que tiene en cuenta algunas particularidades del movimiento de la Tierra alrededor del Sol, y corrige errores de hasta 15 minutos en diferentes épocas del año). Esta y otras sutilezas, pese a ser imprescindibles para un buen constructor de relojes solares, constituyen una complicación innecesaria para los propósitos didácticos.

El dispositivo propuesto utiliza materiales modernos (que, lamentablemente, arruinan el medio ambiente). Esto (lo primero) resulta atractivo para los jóvenes y brinda un uso quizá insospechado para todas esas cajitas de CD que nadie sabe dónde almacenar. Quizás tiene como desventaja el hecho de no ser iguales los ángulos entre las líneas horarias (como sería el caso de un reloj de Sol ecuatorial), algo que merece una explicación. En un dispositivo ecuatorial, cuyo tablero es perpendicular al eje de rotación de la Tierra, queda bien claro que la división se debe a la magnitud del giro sobre su eje que realiza nuestro planeta en 1 hora (360º/24 = 15º).

En el caso de nuestro dispositivo, la justificación de los diferentes ángulos del tablero es más de índole geométrica que astronómica. No obstante, para ayudar a comprender por qué se produce esta deformación, el docente puede utilizar como recurso una analogía con lo que sucede cuando un cartógrafo pretende dibujar en un mapa plano toda la superficie de la Tierra. Dado que el globo terráqueo es (aproximadamente) esférico, podemos preguntar a los alumnos qué creen que sucederá con los ángulos que forman los diferentes meridianos si abriésemos el globo y lo aplastásemos sobre una mesa plana. Evidentemente, esos ángulos se deformarían y las superficies de las regiones polares resultarían desmesuradamente grandes comparadas con las reales. Una deformación análoga ocurre con nuestro tablero del reloj de Sol.

## *Recuadro*

### Cómo ubicar el norte usando las sombras

La brújula permite ubicar la dirección del polo norte magnético. Para latitudes lejanas de los polos indica también, con muy buena aproximación, la del polo norte geográfico o astronómico. Pero esa no es la única manera de ubicar la dirección aproximada de los polos. De hecho, metodológicamente, es una mala manera, pues constituye un método no astronómico de orientación. Hay además ciertas razones didácticas que hacen a la brújula poco recomendable, pues su empleo fortalece ideas previas sobre la supuesta relación entre el campo magnético terrestre y la rotación del planeta. El docente puede entonces preguntar a los alumnos cómo harían para ubicar el norte sin usar una brújula. Sin duda surgirán muchas ideas y quizás las sombras ocupen algún lugar entre estas propuestas.

Una manera de usar las sombras para hallar los puntos cardinales es dibujar una circunferencia sobre un papel y colocar en su centro un gnomon vetical. Si marcáramos con un lápiz sobre la hoja todos los puntos por los que va pasando la sombra de la punta de gnomon, obtendríamos una curva –por lo común una hipérbola– que cortaría a la circunferencia en solo dos puntos. Como podemos imaginar, el primero marcado corresponderá a un momento de la mañana; el otro, a algún momento de la tarde; ambos serán equidistantes del mediodía solar. Si unimos con una recta los dos puntos, tendremos –con muy buena aproximación– la dirección este-oeste. La perpendicular a ella indica la dirección norte-sur.

## *Lecturas sugeridas*

GANGUI A, 2008, 'La precesión de los equinoccios', CIENCIA HOY, 107, octubre-noviembre, pp. 54-63.
ROHR R, 1986, *Les cadrans solaires*, Editions Oberlin, Estrasburgo.

*Figuras*

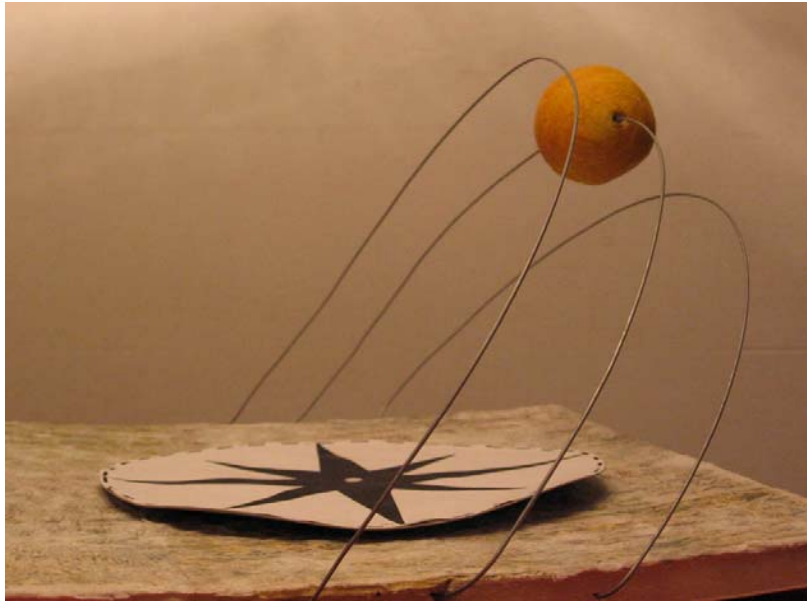

**Figura 1:** Maqueta que representa la trayectoria aparente del Sol por la bóveda celeste en tres momentos del año. Consiste en un alambre semirrigido, doblado con la forma de una semicircunferencia, sujeto a una base de telgopor. Enhebrada en el alambre, se colocó una pequeña esfera de telgopor que indica el Sol. El desplazamiento del Sol a lo largo del alambre permite simular la posición del astro en diferentes momentos del día, en especial en el levante, el poniente y el mediodía solar. Para reflexionar sobre las sombras de los objetos, se puede colocar una varilla vertical en el centro de la rosa de los vientos, a modo de gnomon.

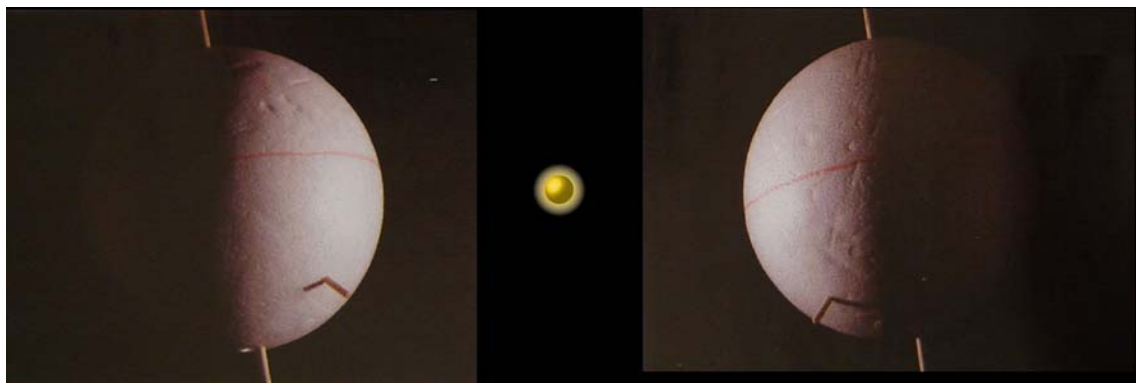

**Figura 2**: Una representación de la Tierra vista desde el espacio (o modelo externo de la Tierra) iluminada por el Sol. La imagen de la izquierda corresponde a la sombra producida por un palillo ubicado en el hemisferio sur durante el verano austral. La imagen de la derecha ilustra lo mismo pero para el invierno austral. La longitud de las sombras nos da indicios sobre el momento del año: a igual hora solar, sombras largas caracterizan los días invernales. El Sol está ubicado en el centro de la imagen con fines puramente ilustrativos. Claramente, esta imagen no respeta los tamaños o distancias reales, es decir, la escala del sistema.

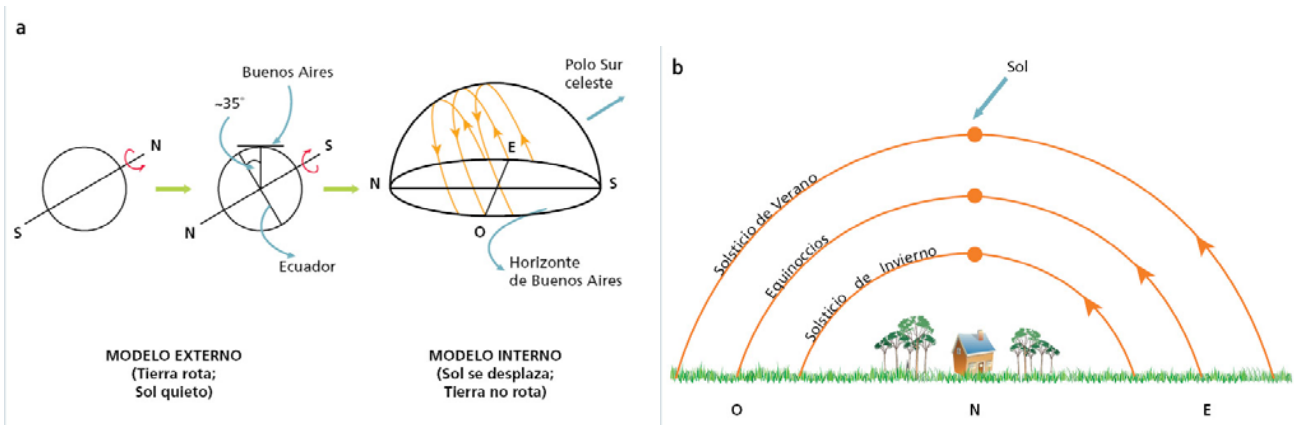

**Figura** 3a y 3b: El esquema de la figura 3a muestra la transición entre un modelo externo y un modelo interno para representar el ciclo día-noche (es decir, el fenómeno visto desde el espacio y visto desde la Tierra). En el modelo externo, la Tierra rota sobre su eje y el Sol se mantiene fijo en el espacio. La orientación del eje de rotación de la Tierra del croquis de la izquierda sigue aproximadamente la representación usual de los globos terráqueos, con el polo norte hacia arriba. En el croquis del medio esta orientación está invertida, y el polo sur está arriba; se señala además la ubicación de un lugar arbitrario del hemisferio sur, como Buenos Aires. Los objetos ubicados por encima del plano horizontal tangente a la esfera son los únicos observables desde esa localidad, porque están por encima del horizonte: las direcciones en el espacio hacia donde se prolonga ese plano constituyen el horizonte del observador. El croquis de la derecha muestra una representación de lo mismo visto desde la Tierra, es decir, en el marco de un modelo interno. Para el observador interior al sistema, la Tierra permanece inalterada y es el Sol el que aparentemente se mueve por el cielo con el correr de las horas del día. El sentido de rotación de la esfera terrestre del segundo dibujo ahora se traduce en un sentido *contrario* para el desplazamiento del Sol por el cielo. En el croquis del medio, la Tierra gira hacia el levante; en el de la derecha, el Sol de aleja del levante y se desplaza hacia el poniente. Esto último también sucede en la figura 3b, una representación que guarda fuerte analogía con la maqueta de la figura 1. Los tres arcos solares representados corresponden a los cuatro momentos especiales de la órbita de la Tierra alrededor del Sol: los dos solsticios (arco largo y arco corto) y los dos equinoccios (ambos representados por el arco del medio, únicos momentos del año en los que el Sol sale exactamente por el este y se pone exactamente por el oeste en toda la Tierra).

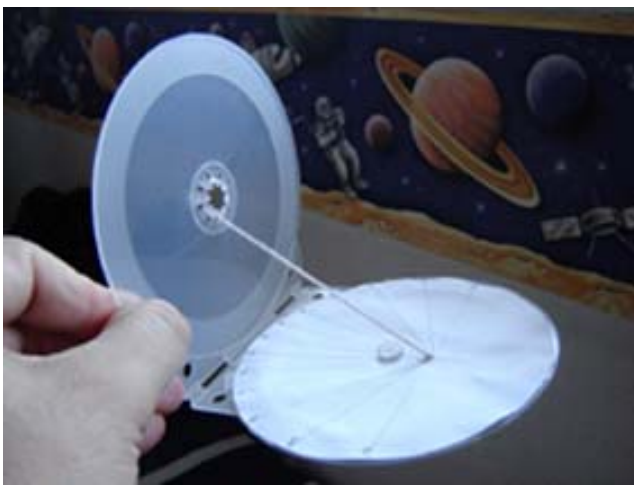
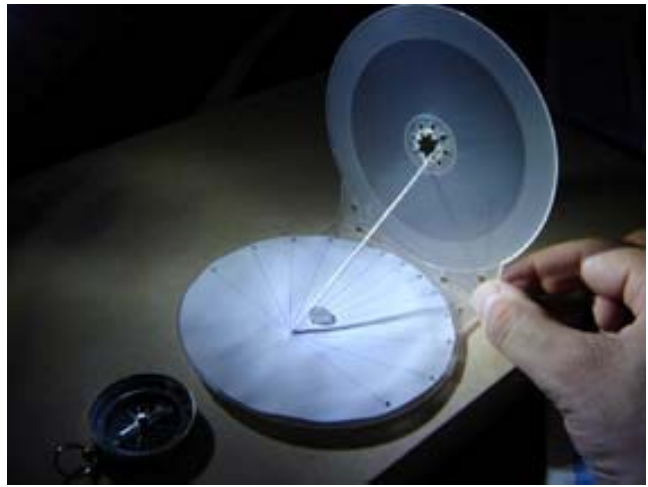

**Figuras** 4a y 4b: Reloj de Sol *compacto*. El gnomon en este dispositivo es un hilo de algodón. Un extremo de este hilo se fija a la base (donde normalmente estaría el CD), un poco por debajo del centro. En ese punto de contacto, se puede realizar una perforación en el plástico (y una en el tablero de papel) y pasar el hilo al otro lado para fijarlo con cinta engomada. El otro extremo del hilo se fija a la tapa del estuche, que debe estar dispuesta en forma perpendicular a la base (para que el dispositivo quede rígido, pueden ubicarse unos ganchos mariposa de modo que cada uno de sus brazos pase por los orificios ubicados en los ángulos de la caja). Queda así formado un triángulo rectángulo cuya diagonal viene dada por el hilo tenso, y los catetos por las paredes de la caja del CD. Recuérdese que la diagonal del triángulo (el hilo que forma el gnomon) es paralela al eje de rotación de la Tierra, y que el Sol, en su movimiento diurno aparente, da una vuelta completa o de 360° alrededor de ese eje en 24 horas. Si el plano del tablero del reloj fuera perpendicular a este eje (y paralelo al ecuador), la división en horas del reloj sería de 15° por cada hora (o sea, 360/24), y en este caso tendríamos un reloj de Sol *ecuatorial*. A diferencia de este caso, en un reloj de Sol *horizontal* como el que contruimos aquí, el tablero no es paralelo al ecuador sino que se apoya sobre la superficie horizontal del lugar en donde lo estamos utilizando. En consecuencia, las líneas horarias no se hallan separadas por 15°, sino que se deforman. El dispositivo mostrado fue construido para la ciudad de Buenos Aires, con una latitud de 35°S (o -35°) aproximadamente. Por consiguiente, el ángulo entre el hilo y la base es de 35° y la dirección del hilo apunta directamente hacia el polo sur celeste, ubicado a 35° sobre el horizonte sur. Para su correcto uso, el reloj debe apoyarse sobre una superficie perfectamente horizontal y orientarse con el hilo en la dirección norte-sur (la meridiana del lugar) y con la tapa del CD ubicada hacia el sur con respecto al tablero.

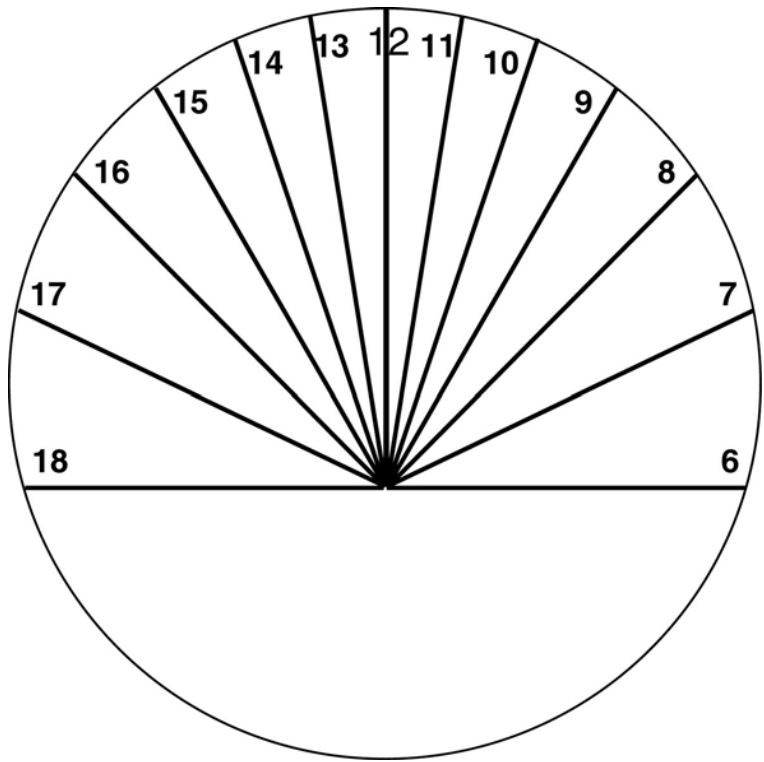

**Figura** 5: Tablero del reloj de Sol horizontal de la figura 4. Para su empleo se sugiere hacer una fotocopia ampliada, de modo tal que el diámetro del tablero sea de unos 12 cm. Esa es la medida adecuada para adaptarse a un estuche de CD estándar. Si la caja fuese de otro tamaño, es necesario adaptar el tablero con la precaución de que el hilo (gnomon) toque la base en el punto del tablero en

que se cruzan todas las líneas horarias. La línea horaria de las 12 debe hacerse coincidir con la dirección norte-sur, con el 12 del lado sur. El tablero mostrado aquí fue diseñado para la ciudad de Buenos Aires, con una latitud de 35ºS (o -35º) aproximadamente. Un simple cambio en el dibujo y uno en la orientación del tablero se hacen necesarios si queremos usarlo tambien en el hemisferio norte, para una latitud de +35º. Identifique el lector esos cambios.

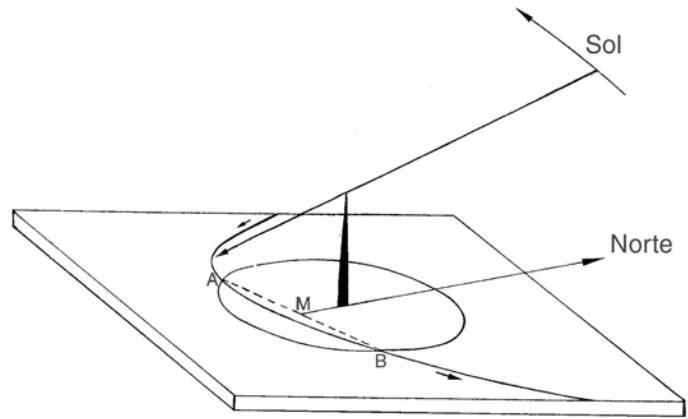

**Figura del recuadro.** Cómo ubicar el norte usando las sombras.

Alejandro Gangui
Doctor en astrofísica, Escuela Internacional de Estudios Avanzados (*International School for Advanced Studies*), Trieste.
Investigador adjunto, Conicet.
Profesor, FCEyN, UBA.
Miembro del Centro de Formación e Investigación en la Enseñanza de las Ciencias, FCEyN, UBA.
*gangui@df.uba.ar*
*cms.iafe.uba.ar/gangui*

María C. Iglesias
Profesora universitaria en ciencias biológicas, FCEyN, UBA.
Docente de enseñanza secundaria.
Ayudante de cátedra, FCEyN, UBA.
*miglesias@cefiec.fcen.uba.ar*

Cynthia P. Quinteros
Estudiante de licenciatura y profesorado en Ciencias Físicas, FCEyN, UBA.
*ufocpq@gmail.com*